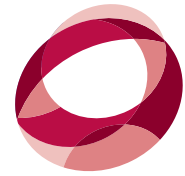

# The Future of Computing Research: Industry-Academic Collaborations

Version 2

Nady Boules, Khari Douglas, Stuart Feldman, Limor Fix (Organizer), Gregory Hager (Organizer), Brent Hailpern (Organizer), Martial Hebert, Dan Lopresti, Beth Mynatt, Chris Rossbach, Helen Wright

IT-driven innovation is an enormous factor in the worldwide economic leadership of the United States. It is larger than finance, construction, or transportation[1], and it employs nearly 6% of the US workforce. The top three companies, as measured by market capitalization, are IT companies – Apple, Google (now Alphabet), and Microsoft. Facebook, a relatively recent entry in the top 10 list by market capitalization has surpassed Walmart, the nation's largest retailer, and the largest employer in the world. The net income of just the top three exceeds $80 billion – roughly 100 times the total budget of the NSF CISE directorate which funds 87% of computing research. In short, the direct return on federal research investments in IT research has been enormously profitable to the nation.

This is just the tip of the iceberg. Although computing-led disruptive innovations tend to dominate the spotlight, computing and data are now integral to nearly every industry. As a result, computing-driven disruptive innovation is taking place across a wide swath of the economy. For example, innovations in the health and medical industries rely heavily on advances in computing power. Agriculture is increasingly automated and there is a tremendous growth in data analytics to improve efficiency, eliminate contamination, and reduce waste – all the way from the farm to the table. In the automotive industry new car models increasingly compete with each other based on the safety, luxury, and automation features enabled by advanced on-board and cloud computing technologies. Service companies, finance companies, retailers, and trading companies increasingly rely on advanced analytics, driven by new sources of data, to improve their operations and compete in the global marketplace.

The central position of computing across these industries is precipitating fundamental changes in academic computing research. For one, *interdisciplinary research* is on the rise. Disciplines such as bio-medical informatics, computational biology, econometrics, robotics, and cyberphysical systems are gaining momentum and showing breakthrough progress. A second change is the *richness and complexity of platforms* and the *concomitant investment in infrastructure* that are necessary for computing research. For example, research involving connected or autonomous cars, smart buildings and cities, cloud computing, the Internet, and manufacturing robotics all require complex, expensive, resource-hungry infrastructure to enable research. Similarly, the recent focus on artificial intelligence and deep learning requires access to a large set of data- and computation-intensive compute nodes to train advanced systems. The third–and perhaps most important–change in academic computing research is the perception that the *time scale of research is shortening.* An increasing amount of research is done with an application in mind. Fundamental or theoretical results are increasingly expected to be complemented by software development, empirical demonstration and statistical validation. At the same time, many universities are encouraging faculty and students to engage in entrepreneurial activities as a way to monetize the intellectual property (IP) that now vests with the University as a result of the Bayh-Dole Act of 1980. It is worth noting that the fraction of PhD in computer science graduates that are going into academic careers, as a fraction of total production is at an historic low (Figure 1).

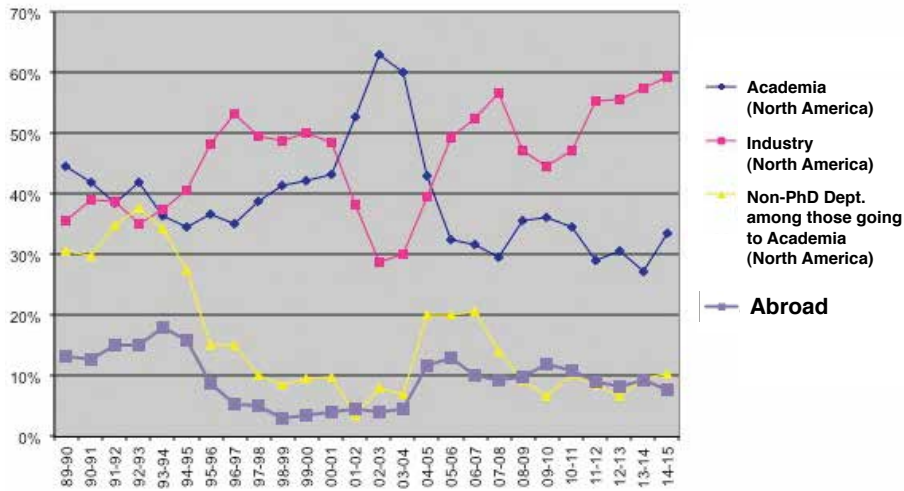

*Figure 1: PhD production and destination from reproduced from the 2014 CRA Taulbee report, Figure 4a.*

The IT industry ecosystem is also evolving. The time from conception to market of successful products has been cut from years to months. Product life cycles are increasingly a year or less, especially when new products are delivered as electronic services, hosted "on the cloud", instead of as installable software or hardware/software appliances. This change has pressured companies to focus industrial R&D on a pipeline or portfolio of technologies that bring immediate, or almost immediate, value to the companies. To defeat the competition and stay ahead of the pack, a company must devote resources to realizing gains that are shorter term, and must remain agile to respond quickly to market changes driven by new technologies, new startups, evolving user experience expectations, and the continuous consumer demand for new and exciting products. We note this is taking place at a time where historically prominent industry R&D labs have downsized or closed entirely, and relatively few new labs are taking their places. This creates a gap between academic research and industry applications, which must be filled in some way.

---

[1] https://www.comptia.org/resources/2015-cyberstates?tracking=resources%2fcyberstates-2015&c=43605

These changes are taking place within a landscape in which federal support for fundamental information technology research is growing slowly, if at all. Further, there are continuing concerns that government programs–both mission and science agencies–are also being pushed toward shorter-term, incremental goals and immediate impact to ensure demonstrable relevance to US competitiveness. Other sources of support for IT research such as direct philanthropic support for computing research continues to play a limited role, with a few notable exceptions (cf. Science Philanthropy Alliance).

Amidst this landscape, the Computing Community Consortium convened a round-table of industry and academic participants to better understand the landscape of industry-academic interaction, and to discuss possible actions that might be taken to enhance those interactions. This discussion was preceded by a survey sent to academics and industry representatives. This survey was designed to provide some current information about the perceptions of the value of academic/industry interaction as well as trends and barriers. This survey is attached as an appendix to this report and is referred to throughout.

The discussions during the round-table, and the data from the survey led to a set of themes that we explore within this report:

1) Is the relationship between industry and academia changing? If so, what drives that change, and how should we respond? Are there long-term risks to these trends?

2) What are current collaboration practices, and how are they evolving?

3) What types of "best practices" could enhance the pace and value of academic research and to accelerate idea and technology transfer? What are the potential barriers?

We close with some recommendations for actions that could expand the lively conversation we experienced at the round-table to a national scale.

## I. The Industry/Academic Landscape

At a high level, the discussions of the current state of the academic/industry ecosystem during the round-table revolved around three "flows" that impact industry/academic interaction: 1) ideas and know-how, 2) people, and 3) resources. Ultimately, new ideas and know-how are what drives innovation, when harnessed to an appropriate commercial opportunity. However, often new ideas can only come into being when the right people and resources come together. Furthermore, much of our fundamental understanding and training occurs in an academic environment, suggesting that a balance between academic and industry people and resources is paramount to keep the innovation system in homeostasis and to support the generation of new ideas and know how.

### *People*

Over the last three years, two new PhDs are going into some type of industry position for every new PhD that goes to academia (Figure 1). Of those two industry positions, roughly one will go into a research position, and the other into some other (most likely development-oriented) position (Figure 2). Looking at the trend data, it is worth noting that this ratio is not as much reflective of a change in the number of students going into industry, but rather a *general downward trend of students going into academia*. Anecdotally, there is a perception among students that working in industry provides the opportunity to have large and immediate impact, larger financial rewards, and to have a "less complicated" existence vis-à-vis academia.

Another recent trend has been a tendency for industry to target academic faculty and, in some cases, entire research groups, to drive specific initiatives. In most cases, this is a reflection of a traditionally academic area of research reaching a level of maturity where it becomes "industry-relevant." Recent examples include computer vision, speech, language, and various learning technologies (such as autonomous vehicles and robotics).

This trend is reinforced by ample examples where a small number of individuals have been able to "move the needle" in major companies, impacting millions of users and thus having large and quite public impact. While this is by no means a completely new phenomena, the scale and frequency (Figure 3) is creating a number of stresses within the academic system as top talent moves to industry.[2]

Interestingly, relatively few of these cases involve major research labs. Indeed, two of the highest valued companies – Apple and Google[3] – do not have delineated research efforts that interact with academia in substantial ways. Yahoo recently closed its research lab.[4] Microsoft is the only one of the highest valued companies that continues to drive a well-known research laboratory, though their mission and scope has evolved to largest technology company as measured by number of employees, continues to support several major research laboratories. Facebook is in the process of creating a research effort; exactly how it evolves remains to be seen. However, compared to the past, investment in industrial research labs seems to be on the decline.

| Table D4a. Detail of Industry Employment | Artificial Intelligence | Computer-Supported Cooperative Work | Databases/ Information Retrieval | Graphics/Visualization | Hardware/Architecture | Human-Computer Interaction | High-Performance Computing | Informatics: Biomedical/ Other Science | Information Assurance/Security | Information Science | Information Systems | Networks | Operating Systems | Programming Languages/ Compilers | Robotics/Vision | Scientific/ Numerical Computing | Social Computing/ Social Informatics | Software Engineering | Theory and Algorithms | Unknown | Other | Total | |
|---|---|---|---|---|---|---|---|---|---|---|---|---|---|---|---|---|---|---|---|---|---|---|---|
| **Inside North America** | | | | | | | | | | | | | | | | | | | | | | | |
| Research | 52 | 0 | 39 | 28 | 29 | 13 | 13 | 11 | 14 | 4 | 5 | 42 | 18 | 15 | 22 | 4 | 4 | 31 | 13 | 23 | 39 | 419 | 46.8% |
| Non-Research | 24 | 0 | 25 | 23 | 13 | 6 | 7 | 15 | 12 | 2 | 16 | 46 | 18 | 13 | 12 | 3 | 9 | 46 | 16 | 18 | 11 | 335 | 37.4% |
| Postdoctorate | 3 | 0 | 1 | 2 | 1 | 0 | 1 | 2 | 0 | 0 | 2 | 1 | 0 | 2 | 4 | 0 | 0 | 0 | 2 | 7 | 0 | 28 | 3.1% |
| Type Not Specified | 6 | 0 | 13 | 4 | 4 | 4 | 6 | 2 | 6 | 1 | 1 | 4 | 5 | 4 | 5 | 2 | 0 | 16 | 5 | 17 | 9 | 114 | 12.7% |
| **Total Inside NA** | 85 | 0 | 78 | 57 | 47 | 23 | 27 | 30 | 32 | 7 | 24 | 93 | 41 | 34 | 43 | 9 | 13 | 93 | 36 | 65 | 59 | 896 | |
| **Outside North America** | | | | | | | | | | | | | | | | | | | | | | | |
| Research | 3 | 0 | 3 | 2 | 2 | 0 | 2 | 0 | 0 | 0 | 0 | 3 | 0 | 0 | 1 | 0 | 0 | 3 | 2 | 5 | 0 | 33 | 61.1% |
| Non-Research | 1 | 0 | 0 | 0 | 0 | 0 | 0 | 0 | 0 | 0 | 0 | 2 | 0 | 0 | 0 | 0 | 0 | 1 | 0 | 0 | 1 | 11 | 20.4% |
| Postdoctorate | 1 | 0 | 0 | 0 | 0 | 0 | 0 | 0 | 0 | 0 | 0 | 0 | 0 | 0 | 0 | 0 | 1 | 0 | 1 | 0 | 0 | 5 | 9.3% |
| Type Not Specified | 1 | 0 | 1 | 0 | 0 | 1 | 0 | 0 | 0 | 0 | 0 | 2 | 0 | 0 | 1 | 0 | 0 | 0 | 0 | 0 | 0 | 5 | 9.3% |
| **Total Outside NA** | 6 | 0 | 4 | 2 | 2 | 1 | 2 | 0 | 0 | 0 | 0 | 7 | 0 | 0 | 2 | 0 | 1 | 4 | 3 | 5 | 1 | 54 | |

*Figure 2: Destination of non-academic PhDs in computer science (from 2014 Computing Research Association (CRA) Taulbee report).*

---

[2] http://www.economist.com/news/business/21695908-silicon-valley-fights-talent-universities-struggle-hold-their

[3] http://cacm.acm.org/magazines/2012/7/151226-googles-hybrid-approach-to-research/fulltext

[4] http://www.businessinsider.com/yahoo-labs-to-integrate-with-product-groups-2016-2

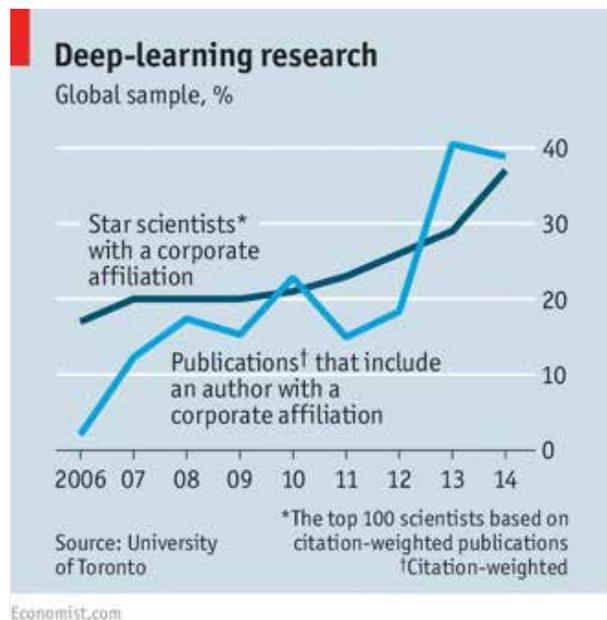

*Figure 3: The impact of the move of top academics in deep learning to industry.*

One of the challenges with this change is the loss of a natural "impedance match" between industry and academia. Industrial research labs typically have an open publishing style, and their employees often attend academic conferences and participate intellectually in the development of their field. This created a natural intellectual flow which, one might hypothesize, reduced the pressure to directly transfer knowledge through hiring.

*Resources*

The value of leveraging industry-centered resources has never been greater. Google, Amazon, and Microsoft have the largest distributed computing operations ever seen, with tremendous resources to expand and innovate throughout the systems stack. However, "resources" are much more than machines and networks – one of the most important resource today (after financial funding) is "access to data". Facebook has the largest set of users in the world, providing unique data resources as well as the opportunity to observe human behavior and to understand trends in socio-technical systems. In application domains, Tesla and Google and many other car companies will now be able to gather unprecedented data on human driving behavior. Intuitive Surgical can observe surgeons at work at the scale of millions of procedures. Large agriculture companies can now observe (and control) equipment, seed, and fertilizer use and resulting crop yields. Every year, the list of unique data and resources grows.

While these opportunities exist, most of these resources are not open to academic researchers. Historically, the academic research created the notion of open-source, which in turn created a new vehicle for academic-industry collaboration. However, the data and resources generated by industry are not (and likely cannot be) open, making collaboration around these resources difficult.

## 2. Collaboration mechanisms – Opportunities and Challenges

Industry and academia are already strongly intertwined. For example, nearly ¾ of the survey respondents indicated they have some type of industry sponsorship, half indicated paid consulting arrangements, and 95% indicated industry-hosted interns in some form. Even allowing for possible sample bias, there is clearly a vibrant exchange between industry and academia. However, based on discussions at the round table. It is also clear that no single collaboration template is either possible or desirable because of the wide variance in the type of research (e.g., basic to near product integration), goals, and size of the projects. In what follows, we provide a coarse mapping of the space, with the caveat that every relationship will undoubtedly have its own unique character and nuances.

It is also important to differentiate the goal or objectives of the collaboration from the mechanisms that are used to achieve them. As noted above, there are three dominant goals or outcomes of an industry-academic collaboration:

1) *ideas* with actionable IP, such as algorithms, designs/architectures, open source software, or new research directions;

2) *resources, data, things or services,* when the product of the collaboration takes the form of software or hardware artifacts or data moving between academia and industry

3) *people*, when the main objective of the collaboration is the transfer of people, research and students, with specific skills to industry, or for the creation of an ecosystem (of developers, of users, etc.).

The details of the collaboration mechanism depend on the mix of desired outcomes. It is also important to note that there are other goals – for example, collaborations might be designed to enhance educational opportunities for students or employees, or to foster a broader strategic relationship.

Below we describe three common collaboration mechanisms used in industry-academic partnerships, informed by an understanding of the challenges that can affect them. For each of them we indicate pros and cons and discuss the current difficulties and challenges in implementing them, from both perspectives.

### *Contracting*

This is a common collaboration model, wherein a contract or grant is established from a company to an academic institution with a specific statement of work and deliverables. This model is standard way to connect industrial development with academia, though less common for an industrial research team. The advantage of this mechanism is that it is well established and most organizations are equipped to make use of it – there are long-standing terms and conditions templates and expert staff at every research university for the negotiation and implementation of these contracts.

While this mechanism has been the bread and butter of industrial development-academic collaborations, it now faces considerable challenges in a rapidly moving tech ecosystem that operates more like a startup then an established industry. Most would argue that, by itself, research contracting is no longer sufficient for several reasons:

- *Timescale mismatch.* As previously discussed, the industry timescale tends to be considerably shorter than that of academia. It is often difficult to justify long term, multi-year research projects. This is particularly the case in the fast-moving frontier of the tech industry where products are rolled out in months rather than years. Industry goals for products can also change with no notice.

- *Project granularity.* Due to timescale mismatch, short-duration projects, (e.g., 6 months, deliverable-heavy projects) are very disruptive to academic environments because it requires stability of student and staff investments. As a result, such short-term projects most often produce what is already available with limited innovation. It is important to note that this issue may deepen the gap between major research universities, which are able to put in place the broader and more flexible mechanisms described later, and other universities, which may have to rely more on short-term efforts to the detriment of their long-term research capacity. Even in six-month projects, an industrial "agile" project will have a constant cadence of team meetings and milestones: if not carefully managed, they will strangle any chance for research innovation. Also, such a cadence, unless very well managed, will squeeze out any time for publication due to the difficulty in justifying the "extra" experiments needed for scholarly acceptance of research results.

- *Lack of Capability Differentiation:* The contracting model is most appropriate when the industry entity has little to no in-house technical capabilities in the technical area of interest. In an increasing number of cases, the industry entity has in fact had significant internal resources and the value of the university research given the IP and T&C complications becomes far less attractive. In this case, a mechanism in which the industry's existing resources become more integrated with the university's becomes more attractive motivating the shared entity model described below.

◗ *Skills vs. IP:* It is increasingly the case that the industry partner is more interested in building up technical skills internally, than acquiring technology and IP. Single inventions in computing tend to have low individual value, since a complex device or system may embody thousands of patents, with each one contributing a small amount to the final value. Regulatory policies such as Bayh-Dole and work-for-hire limits can create barriers to IP or opens up the risk that IP could be "resold" later to competitors. Universities have used different ways of trading off academic interests and regulatory compliance with industry needs in terms of IP. For example, exclusive licenses in field of use or restrictive clauses that has to be agreed to by individual PIs.

◗ *IP.* The toughest issue in most university-corporate interactions is IP ownership and control, especially when one of the parties does not understand the true value of the IP to be produced. Often, there is considerable tension between the university's IP office and the corporation's lawyers, and this may override the eagerness of scientists who want to interact. It is not rare for it to take more time to negotiate the IP terms than the length of the proposed project or sabbatical. This negotiation time can be massively detrimental to establishing a partnership. In fact, some companies have internal rules that call for the company to abandon negotiations if they cannot close the agreement in a fixed number of months. Problems also arise from the differing incentive between the two. The university lawyers worry they will be seen as having given away a huge amount as a result of a contract, but there is little penalty for blocking one. Companies tend to view IP rights as a business decision about the expected royalty stream or the value of the freedom of action. Startups–which can emerge as a result of collaboration–frequently have a key piece of intellectual property that justifies their funding and companies do not enjoy helping their competitors through leakage of their own IP.

### Industrial Gifts/Grants/Fellowships/Internships

When the relationship between academia and industry is through an industrial research lab, the most effective mechanism is usually some form of gift or unrestricted grant. A research lab has the long-term time horizon that can focus on supporting an academic or their students in an area of interest to the parent company. As long as technical results and good students are produced (or if a vital ecosystem is created that is of value to the corporation), the lack of formal deliverables and defined milestones can be supported. This requires maturity on the part of the industrial partner, since all the important results will be published. Hence they must plan to jump quickly on innovations, or have a model where improvement in a subfield will produce a "return on investment" to the parent company even if the company does not exploit the specific technology. Although there is no data (of which we are aware), the perception is that the number and size of such "unrestricted" gifts have declined as the number, time horizon, and size of industrial research labs have declined.

### Direct Skill Transfer

Contracting implicitly presumes there is a "work product" that the industry partner can clearly describe and that a university team can supply. However, in many cases the skills of the personnel involved in the collaboration are more valuable than the immediate research product themselves. Hence, it is natural that, in some cases, the collaboration mechanism reduces to transferring personnel from academia to industry.[5] In a sense this is the extreme opposite of the contracting mechanism: rather than paying an external person with the necessary skills to do the work, the company acquires the skills to do the work internally. From industry's point of view, it is a particularly effective way to quickly establish a position in a new area. It bypasses many of the complexities of contracting – fewer IP issues; better agility with respect to project goals and timelines (which bypasses the timescale issue), and direct team integration. From academia's point of view, it can be a good way to increase recognition and to receive revenues from IP transferred to industry in the short term. While short-term, isolated interactions of this type may be mutually beneficial, sustained skill transfer in any given technical area may not be, especially because of the risk of compromising the training and research capabilities of academia, which produced these skills in the first place. While there are many examples of such interactions, there is no generally accepted set of practices. Collecting case data, as permitted given the confidentiality limitations, would be valuable.

---

[5] Note this mode of transfer is far from new – see the "Evolution of Lisp" by Steele and Gabriel: http://dx.doi.org/10.1145/234286.1057818

*Shared Entities*

Lying between the two extremes of contracting and skill transfer, shared entities are a compromise position that combines the internal resources from industry with the research resources from academia. In effect, it is a strategic merger in a "neutral territory" that provides strategic focus and agility but preserves many aspects of the academic environment. Put another way, it is a new form of industrial research lab, but one that is outside the legal boundaries of the company. Shared entities can be attractive to industry because they are co-investing and thus are using internal resources more effectively. Additionally, it addresses the timescale issue by incentivizing industry to engage in longer relationships– generally through master agreements–instead of individual project-based agreements.

Shared entities may take different forms: on-site labs sharing personnel from industry and academia; industry personnel embedded in university labs; and university personnel, both faculty and students, embedded in industry. Unlike the contracting mechanism, there is no standard template or recorded best practices. Like with the skill transfer, it would be extremely beneficial to collect information that will recommend best practices to facilitate this type of mechanism. In particular, two classes of challenges need to be addressed in this mode of collaboration:

◗ *IP:* Because the work is conducted jointly, creative approaches to IP are necessary – for example some form of joint IP and/or prenegotiated license structure. Defining the IP terms for a long-term agreement is difficult and so there is a need for a continuing structured review of existing regulation and agreements to facilitate IP for shared entities. This is of course very challenging because there is variability across different cases.

◗ *Academic practices:* Shared entities require flexibility on the part of the university partners. This could come in the form of part-time leaves of absence for faculty (or students, or research staff) to work more closely with the industry partners for example. Such practices are often difficult to implement or not allowed under standard university practices. It is imperative that these practices evolve to allow these industry collaboration mechanisms.

*Community or Consortium Model*

The community model involves sharing research among a community of industry subscribers (e.g. as a consortium). This model can be an attractive way of taking advantage of open-sourcing as it allows all partners to contribute to a single shared resources rather than developing it each independently. It bypasses many of the issues associated with the other models but requires a higher degree of sharing on the part of the industry partners. A closely related model is where a single industrial sponsor supports an ecosystem of academic researchers to build new open-source software and related curricula on a common open-source foundation. For example, using this model (in the mid 2000s), IBM supported the Eclipse.org platform through a series of Eclipse Innovation Grants ($10-30K), which funded new open-source software development, as well as creation of Eclipse-related curricula.

Within computing, building communities around software has an established history of open-source-based sharing that originated within the academic community. A variety of well-established licensing models exist, facilitating transfer or share of intellectual property. Further, code is an artifact that can be modified and manipulated to improve or customize functionality, providing a way to produce "value-added" variants, even if large portions of the code base is shared.

Today, communities form around resources other than software. In particular, many researchers are at least as interested in access to the data companies hold as they are in receiving direct support for research. Frequently, relevant corporate data is seen as the "crown jewels" and is of great competitive value. Consequently, the risks of data (or data product) leakage through research collaborations may offset the perceived benefits of collaboration. Proper attribution in research papers and a need or desire to anonymize the data are also key issues (indeed, these issues are also in flux in academia as well). Data that includes personally identifiable information (PII) or which could expose trade secrets are critical to protect and companies justifiably wish to avoid the potential liability associated with privacy or competitive risks. Although anonymization is a possible solution, it is expensive to sanitize or depersonalize the data and so management may not see the value in investing in it. Similarly, some research partnerships have tried creating synthetic data based on real private data (sharing common statistical properties), but there are no community standards and best practices on when that approach is valid.

Other communities have or are anticipated to develop around platforms that are capital intensive and thus will only exist within a few entities – cloud computing, social platforms, vehicle technologies, smart grids, and so forth. Each of these new communities will be an opportunity to create a synergistic community-based collaboration between industry and academia, but each will present its own unique challenges.

## 3. Best Practices for Research Collaborations between Academia and Industry

Given this evolving landscape of interactions, it would be presumptuous to expect that we could predict the best mechanisms to support, or create fixed models for industry-academia collaboration. However, there is a growing pool of expertise and experience that could be collected to help inform future efforts.

*Focus on Concrete and Grounded Collaborations:* Ab initio deals closed at very high levels rarely survive or prosper. A CEO may have a photo-op with a university president and promise significant funding and long-term collaborations. On the company side, the responsibility for keeping the relationship going will devolve to lower levels in the corporation over time; so too will the budget responsibility. This generally leads to narrowed focus, slow decay of funding, decreased commitment for the whole relationship, and difficulty getting individual scientists and engineers to actively participate. On the university side, the administration cannot order faculty to do anything, and the good will between the faculty and administration at the start of the collaboration will decline with increasing numbers of meetings and decreasing breadth of interest.

Positive examples often involve a local university of particular overall value to the company (i.e., relevant specialties, lots of students, faculty in multiple departments interested in its problems) or a physical laboratory on or adjacent to the campus. Proximity and synergy means that collaborations will evolve on their own at some level from interplay between researchers at active centers in the university and specific projects or departments in a company. This sets up the possibility of long-term interactions, framing of interest problems on the corporate side, and deep context learning on the university side. There will be fewer photo-ops, but more papers, corporate impact and cash flow. It is essential, however, for both sides to understand what the other side is getting out of the collaboration. How will both parties measure success or failure, both on a short-term basis and when the agreement comes up for renewal?

*Establish Sustained and Embedded Interactions:* As discussed previously, establishing and advancing a successful interaction can be very difficult and involves establishing common interests and trust on both sides. A mechanism that has worked very well uses internships or sabbaticals to place those on the university side in industry, with a complementary embedding of corporate technical people at the university. This results in people who acutely understand the actual problems being faced, not just oversimplified versions of them, and real techno-social interactions that lead to trust and understanding and actual contributions. It is important to distinguish between the relationships an academic institution might have with the research arms of a corporation from those it might have with the engineering or development groups. Fewer companies have identifiable research organizations than engineering or development groups but those that do increasingly expect some clear value creation in the company on moderate time scales (less than 3 years). If the industrial partner has no in-house research organization, then a constant education of the industrial management of how to measure and evaluate the partnership is absolutely necessary. Additionally, the research organizations themselves often struggle for continued relevance and contribution and may be in need of some quick hits from their own collaborations leading to competition with academic consultants to the engineering group.

*Create Reusable IP Transfer Vehicles:* One effective technique to set up IP agreements is to painstakingly craft a master agreement. This works well when there are expected to be many interactions between a company and a university. A master agreement creates a natural "corridor" where new activities then only require a quick and easy addition to the master agreement. The details of these agreements are often confidential, making sharing of best-practices difficult. There have been attempts to write boilerplate agreements (especially in the context of open source creation and open research collaborations – see next section) that have been applied broadly on a national scale. These have had limited success, but it may be time to try again with a constellation of corporate and university lawyers in the context of both open software and open data.

Open source can be a way around IP and sharing problems. Putting research results into the open literature provides a counterweight to the growing focus on creating University-assigned IP due to Bayh-Dole. If there is a promise that results of collaboration will be openly accessible, this may quiet concerns about the company losing value from the interaction; although it does raise the risk of helping the competition. Hence it is essential that the industrial partner know how to measure ROI from a growth in the relevant open source ecosystem, and not be surprised when the innovation shows up in the community. The same applies to making detailed data from the research Open Access, as well as other infrastructure.

*Create Models for Sharing Resources:* It is readily acknowledged that open sharing of data accelerates innovation and discovery. In the biomedical sciences, NIH has recently taken a strong stance toward supporting data sharing. The computing domain needs to follow suit and share more data among industry and academic partners. Protection and confidentiality issues for industry data need to be addressed, starting with recording best practices in existing successful agreements. In the other direction, universities have useful data to share but not necessarily the resources to maintain and share it. Industry could play an important role in participating in common data resources.

*Include Education:* The discussion so far has focused on research. However, as industry needs skills and talents often more than technology and IP, the question arises – should there be more direct involvement of industry in the education and training functions of academia? There is indirect involvement through funding of basic research, which contributes to student training, but are there other, more direct mechanisms? Small-scale examples include sponsoring of capstone projects or professional programs. If such mechanisms would be feasible, industry would be engaged in the process of producing the skills that they need. Another possibility would be to bring industry professionals in to teach mid-level courses as adjunct faculty on a regular basis.

# 4. Conclusions and Recommendations

Technology provides a path to the future, and computing is increasingly at the heart of many new technologies. Human-centered computing, big data analytics, extensive machine learning, computing with a societal application, and increased interaction with the physical world are all a part of this new paradigm.

Taking advantage of future possibilities will require a balanced national portfolio that includes both long-term and basic research in computing – the kind that is fundamental to future innovation – as well as more application-driven and applied research. Putting in place mechanisms for linking research to innovation and commercialization and will only grow in importance for our national innovation cycle. Just as we cannot depend on industry to do the fundamental research, we cannot expect academia to grow to fill the applied research gap without additional support mechanisms to do so. Industry-academic collaborations thus offer a mutually beneficial way to support long term, fundamental research, to translate research ideas to industry-specific needs, and to satisfy the need for highly trained students who can build new innovative tools and products.

In reflecting on the results of the survey and the round-table discussions, below are some concrete actions that could be taken to enhance the future vitality and impact of academic-industry interactions:

1) Establish a means of measuring and benchmarking industry/academic interactions. It is hard to assess or improve something that cannot be measured. We know surprisingly little about what sorts of flows – people, resources, or ideas – currently exist between academic institutions and companies. Some aspects are relatively easy to measure – for example the Taulbee Survey already measures the flow of PhDs to industry. Some aspects are in principle measureable – for example, most universities have some way of tallying direct industry research support to faculty. However, many other aspects of industry/university interactions – e.g. funding for academic sabbaticals in industry or in-kind contributions – are hard to measure. Perhaps there are ways to begin to tally these flows.

   Create a repository of best-practices for industry/university interactions. Too often, researchers or companies "re-invent the wheel" by recreating organizational structures, legal frameworks, or term sheets that exist in other areas. It is not uncommon that a collaboration stalls out because of legal considerations – the survey results point to IP barriers as the most frequent limitation on interaction. It is interesting to note that the academia-industry survey specifically points to people-oriented mechanisms as being of most interest; perhaps creating models for those flows would be a place to start.

2) Recognize that there is a need for career paths that may combine elements of a traditional academic career in a university research and education setting with career paths that involve significant time within a new or established company, and create mechanisms that support such career paths. Examples would include sabbatical support for industry research staff in academia, personnel loan arrangements that allow academics to work in industry for a limited time but retain their academic position and seniority, and so forth.

3) Consider ways that advanced infrastructure can be made widely available to the research community. Currently, some universities are able to build their own advanced infrastructure; others depend on collaborative relationships with industry to gain access to commercial-class platforms and data. However, not all investigators have these opportunities and thus cannot participate in these areas of research. Finding ways to make advanced computing and devices, large data sets, and unique facilities more widely available will benefit industry (it will create "power-users" for their infrastructure), academic research (avoiding wasted time and resources replicating capabilities already in existence), and education (students will learn on the latest and greatest).

4) Convene a long-term forum or body around industry-academic interaction. A key value-point here is the fact that many non-traditional industries are growing computing-related research groups. Creating a mechanism that allows these groups to become visible to prospective problem solvers and employees could create a driver to ensure such a forum is well-attended and continues to maintain value and energy. An alternative would be to convene workshops or conference tracks within specific areas of interest, thus providing a more distributed and area-focused means of conversation.

While we cannot predict where the future will go in detail, we know that technology will continue to play a large (and most likely increasing) role in our national well-being. The current boom in innovation is based on many years of fundamental academic research that produced accumulated technological achievements. This knowledge and results have been transferred to industry via many mechanisms, but the dominant force is people; that is, students graduating and joining the work force. Academia is building not only the inventors of tomorrow but also the market (end-users) of tomorrow, thus, creating the demand for new and exciting products. Collaborations between academia and industry will continue to play a central role in the transfer of long-term and fundamental research into the US economy. Recognizing and supporting this transfer will provide mutual benefits to all stakeholders.

## Appendix 1: Identifying Future Opportunities for Industry/Academic Interaction: Two Case Studies.

During the round table, one exercise was to identify areas where stronger interactions between academia and industry would have an impact. From these discussions, two examples, one in core computing and one in applications of computing, emerged. These examples are intended to be illustrative; there are many similar examples where intimate interaction between IT/CS research and industry are needed.

### Computing and Devices: The Automotive Industry

The automotive industry is undergoing a technological revolution. With exponentially increasing electronic (and software) content and interconnected embedded systems cars are becoming huge, complex distributed computer systems on wheels. There are over 200 Electronic Control Units (ECUs) and 100 million lines of code in a modern luxury car. By comparison, there are "only" 5.7 million lines in an F-35 fighter aircraft. These new systems are much more complex than the relatively simple stand-alone computing systems that once controlled basic engine and chassis functions and are evolving to become one of the most sophisticated, widely distributed cyber-physical systems that exist. They represent a class of systems that are characterized by:

◗ *Deep physical interactions*
◗ *Deeply embedded electronics*
◗ *High degrees of computation*
◗ *Rich needs to communicate*
◗ *Pervasive integrations (cyber and physical)*
◗ *Highly coupled with human (driver) behavior*

These changes present designers with major challenges that demand the attention of the computing community.

The role of computing will extend well beyond the individual car. One way to manage the increasing computerization of cars is to establish an Intelligent Transportation System (ITS). Intelligent Transportation Systems are "advanced applications which… aim to provide innovative services relating to different modes of transport and traffic management and enable various users to be better informed and make safer, more coordinated, and smarter use of transport networks."[6] The potential benefits of ITS are huge: enhanced roadway safety, real-time traffic management, improved thoroughfare, enhanced energy efficiency, and reduced emissions. In order for society to reap these benefits, we must anticipate and support efforts that are considered foundational for Intelligent Transportation Systems. The challenges fall into several broad groups that speak to a broad spectrum of computing-related disciplines, including cyber security, management and verification of complex software and hardware systems, trustworthy and reliable computation, and the training and educating of current and future workforce in the technologies of cyber physical systems.

Reliability of computation, and, by extension, safety is also important. Advanced control strategies and architectures are needed to ensure "fail-soft" and fail-operational", required for semi-autonomous and autonomous driving. To achieve the control accuracy and reliability required for advanced active safety and autonomous driving systems there is an increasing need for more reliable sensors, communications, actuators, and computational methods (that are able to handle unreliability) than is available today at affordable cost. Diagnostics and prognostics of CPS systems present a challenge due to their complexity, but at the same time they are considered key enablers for systems service and repair and customer peace of mind. CPS systems are also challenging the current workforce since complexity generally increases faster than capability. Therefore, we need to continuously upgrade our workforce, the ones already working as well as those who will be entering the work force. These challenges require an intensive effort on the part of all stakeholders, OEMs, suppliers, in cooperation with academia and governmental research institutions.

---

[6] https://en.wikipedia.org/wiki/Intelligent_transportation_system

Another challenge that must be addressed is the management of software and hardware complexity. The structure of operating systems remains a huge barrier, as software systems typically need to be redesigned to accommodate the architectural diversity of new hardware. By using today's operating systems one ends up with a dangerous house of cards. Additionally, components and subsystems can no longer be designed and developed in isolation and then integrated into the vehicle. Now, complete systems have to be integrated at the outset of the design process and in setting system requirements to comprehend mutual interactions at deeper and deeper levels. This will require new system engineering and design tools for integration into the vehicle, new development processes, new processes for the integration of manufacturing plants and supply chain. On top of this the integration of sophisticated control algorithms involving a large number of code lines makes it increasingly difficult to verify and validate using conventional manual approaches. The use of emerging, systematic "Formal Methods" techniques are becoming essential for the design of reliable software.

### *Computing in Large Scale Heterogeneous Systems: Operating Systems*

In the previous section, it was pointed out that current operating systems (OS) design is a limiting factor for the development of complex cyber physical systems like cars. However, the structural problems in modern operating systems are not limited to the automotive industry. In a nutshell, a critical impending challenge to computing is the poor fit between the post-Moore's law hardware platforms and the structure of abstraction layers in modern system software like operating systems and hypervisors. Emerging platforms will almost certainly be heterogeneous and distributed, and will also incorporate parallelism and concurrency as crosscutting concerns. This is clearly among the most critical upcoming challenges for computing at large, regardless of who does or does not collaborate to address it, but there are aspects of this problem that make it particularly well-suited for collaboration between industry and academia.

First, we should examine the problems inherent to the current status quo. We are reaching the limits of Moore's Law and Dennard scaling. What are the implications of their demise? The performance and efficiency gains in future platforms will be achieved largely through specialization – algorithmic, architectural, or both – and distribution. The dominant impact of specialization will be in the form of architectural heterogeneity (e.g., GPUs, FPGAs, crypto processors, image co-processors, etc.). Broadly speaking, specialization and distribution will move computations to the resources best suited to perform them whenever it is profitable under a given goodness metric to do so. Moving data to GPUs to accelerate parallel compute phases, or performing work initiated by a mobile device in the cloud are common illustrations of this pattern. The important observation is that in the future, the need to use specialized resources in common-case programs fundamentally means programmers must cope not just with heterogeneity, but with all the challenges of distributed computing, including the thorny ones like concurrency, fault-tolerance, and consistency that continue to fascinate the systems research community to this day. Supposedly "modern" system software like OSes and hypervisors are designed with a goal of hiding these complexities and providing a uniform abstraction of computing fabric to programs; one which is by design independent of the physical hardware. To first order, this has been accomplished by de-coupling concerns such as heterogeneity, failure, concurrency, and distribution.

However, it is no longer tenable to treat these issues as separate concerns and the convergence of these concerns implies there are some hard problems that have to be solved.  If common-case programs must use specialized architectural features to gain reasonable performance or power efficiency, a monumental programmability problem has to be eliminated because these devices are challenging to program, especially when they must collaborate with conventional processors and other specialize computer engines. There has been progress with programming GPUs for an interesting set of applications. But that is a solution for just one type of compute device in isolation. This approach will not scale if we are going to have to repeat that effort every time a new accelerator becomes available?  If common-case programs must use distributed resources, a similar programmability problem must be addressed. Front-end programming for "cloud computing" has certainly enjoyed progress, but at the end of the day, the systems community is still struggling with fundamental tradeoffs between performance, consistency, and programmability.  While this effort has certainly yielded a wealth of cryptically-named, difficult to reason about forms of "consistency", there is a lot of uncharted territory here – for example the needs that will emerge as technologies such as neuromorphic computing and application domains like virtual reality and augmented reality start to enter the cloud ecosystem. Programming for distributed computing is far from solved. Perhaps more importantly, distribution implies major challenges around privacy and security.

This brings us to our main question: why is this a good area for industry-academia collaboration? Many recognize this area as a problem but are unable to deal with it because it requires radical changes in system layers for which the financial incentive to change is too distant. Too many things depend on various facets of the current structure. Restructuring either impacts existing critical programs or is simply off the table for ROI reasons, even if it is obviously necessary for the long term. Academia is better positioned than industry to take the kind of radical positions that are going to be required. Proposing system structures and abstractions that leave legacy code to die is unattractive no matter which you are, but it is tenable in an academic setting.  On the other hand, radical change at the lowest layers of the software stack entails a high ratio of engineering effort to research result, making such lines of inquiry unattractive to many academics. Collaboration on these topics between academia and industry may enable research that mitigates the risks to both, while leveraging the strengths of each environment.

## Appendix 2: CCC Industry and Academia Survey

In spring 2015, the CRA and the CCC released two short surveys, one for the academic community and the other for industry, to learn about academic-industry interactions. The purpose was to provide a picture of the types of interactions currently taking place, and to identify common barriers to those interactions. In addition, the CRA and CCC were looking for feedback on ways that they could strengthen the relationship between the two.

The first set of questions in both surveys were basic background questions asking for organization name, job title, and contact information (if respondent wanted to be contacted). Survey participants were asked to identify their role in their organization (e.g., staff researcher, department manager, department chair) and respond from that perspective. The CRA and CCC were seeking a broad representation of managers and researchers.

The questions in the second part of the survey differed depending on whether the survey was geared toward academia or industry. The academia survey had a total of 13 questions and the industry survey had a total of 17 questions. The entire survey was a qualitative effort to gain insight into academia/industry interaction.

### Academic Survey

The academic survey was sent out to 213 academics, which included mostly computer science department chairs. There were 60 total responses, which is a response rate of about 28%. The majority of the respondents were from public institutions (75%), not private institutions. Respondent's organization varied greatly in size and type, from 20 faculty members and 500 students to 900 faculty members to 20,000 students. Some of the respondent's organizations had undergraduate students only, while others had undergraduate and graduate students. Finally, respondents were asked to identify their role in their organization. A majority of the respondents were the department head / chair (77%). The rest of the respondents were a mix of Dean and Professors.

## Academic Survey Results

The first academic question asked, what are the types of interactions you have with industry? Respondents were asked to select all that apply.

The majority of the respondents said that industry hires their undergraduates and hosts interns from them. Other types of exchanges that were noted include, distinguished lectures, individuals in industry hired as adjunct faculty members, and collaboration on undergraduate capstone projects (Table 1).

The next set of questions asked what barriers do you commonly encounter and are hardest to solve when working with industry? The majority of the respondents said that intellectual property and finding a good contact person within industry are the most commonly encountered barriers and are also the hardest to solve. Other barriers that were noted include, making the right connections and finding the right pitch for doing research within industry. The respondents noted that industry often wants numbers ("We reached n thousand students and x hundred teachers!") to promote their product, while academics themselves are more interested in insight for research.

The last question asked academics to identify opportunities that they believed would be most effective to improve the connections between academia and industry. Respondents were asked to select at most three.

The majority of the respondents said that providing better methods for interaction/exchange of personnel between academia and industry would be the best way to improve the connection. Creating better vehicles for exposing and engaging academic research programs with industry would also be an effective way to improve the connection (Table 2). Other ways to improve the connection is working through large government grants that require industry involvement but also require academics doing 'further out' research.

| Answer | Response | % |
| --- | --- | --- |
| Hire our PhDs and post-docs | 33 | 75% |
| Hire our undergraduates | 42 | 95% |
| Host interns from us | 42 | 95% |
| Work on collaborative projects funded by industry | 33 | 75% |
| Work on collaborative projects funded by a third party (e.g., DARPA) | 24 | 55% |
| Paid consulting arrangements | 24 | 55% |
| Tech transfer of research results | 27 | 61% |
| Access to data or infrastructure to evaluate research ideas | 19 | 43% |
| Recruiting mid-career faculty from industrial labs | 11 | 25% |
| Hosted for a sabbatical | 17 | 39% |
| Other types of exchange (please specify): | 10 | 23% |
| Work on collaborative projects without funding | 20 | 45% |

*Table 1. Type of interactions academia has with industry.*

## Industry Survey

The industry survey was sent out to 18 individuals in industry with instructions to forward to colleagues. The exact number of individuals who received the survey is unknown. A total of 66 surveys were completed. The majority of the respondents who filled out the survey were from IBM Research (38%). Another common company was Intel (12%). The others were a mix of large and small companies like Yahoo Labs, Microsoft Research, Big Switch Networks, Corsa, and Snapchat. Respondents were asked to approximate the size of the organization that they managed. The numbers ranged in size from 5 individuals to 300. Finally, respondents were asked to identify their role in their organization. A majority of the respondents were research staff members (45%). Other respondents included directors and lab managers. A majority of the respondents were in the industry basic or applied research area (77%).

## Industry Survey Results

The first industry specific question asked, what types of interactions do you have with academic researchers? Respondents were asked to select all that apply.

The majority of the respondents said that they host graduate student interns as well as hire PhDs as permanent staff members. Other types of exchanges that were noted include, issuing awards and providing gifts to universities (Table 3).

The next questions asked, what barriers do you commonly encounter and are hardest to solve when working with academics? The majority of the respondents said that intellectual property is the most commonly encountered and is also the hardest to solve. One respondent said that intellectual property becomes an institutional issue on both sides and there may be little room to maneuver. Other barriers that were noted include, nondisclosure agreements and open source/open access vs. IP protection.

The next question asked industry to identify opportunities that they believed would be most effective in improving the connections between academia and industry. Respondents were asked to select as most three.

The majority of the respondents said that better training of students for work in an industrial setting (e.g. professional programming, working effectively in teams, awareness of new technologies, good communication skills, etc.) would be the best way to improve the connection. Providing better methods for interaction of personnel between academia and industry and creating better mechanisms for shared data or infrastructure would also be effective ways to improve the connections. Respondents also mentioned that bringing students for extended stays to industry and creating a form where important technical issues that academic students and researchers may not be aware of could be presented and discussed (Table 4).

| Answer | Response | % |
|---|---|---|
| a. People-oriented — e.g., providing better methods or best practices for interaction/exchange of personnel between academia and industry | 33 | 79% |
| b. Process-oriented — e.g., creating better vehicles for exposing/engaging academic research programs with industry | 27 | 64% |
| c. Resource-related --- e.g., creating better mechanisms for shared data or infrastructure. | 18 | 43% |
| d. Communication-related — e.g., creating a clearing house for CS PhDs who would be interested in summer internships at a company. | 12 | 29% |
| e. More opportunities for people working in industry to attend, speak at, or publish at research conferences and journals?  (e.g., industry tracks at conferences, conferences located near major cities or industrial hubs, survey papers or panels on major trends or technologies in industry, etc.) | 10 | 24% |

*Table 2. Most effective ways to improve the connection between academia and industry.*

| Answer | | Response | % |
|---|---|---|---|
| Hire PhDs as permanent staff members | | 34 | 87% |
| Hire PhDs as temporary staff (e.g., limited term postdocs) | | 22 | 56% |
| Host graduate student interns | | 35 | 90% |
| Work on collaborative projects funded by your organization | | 28 | 72% |
| Work on collaborative projects funded by a third party (e.g., DARPA) | | 15 | 38% |
| Provide research funding to university faculty | | 20 | 51% |
| Hire academics as consultants | | 14 | 36% |
| Hosting visiting professors (e.g., sabbaticals) | | 24 | 62% |
| Issue awards for promising early career faculty | | 12 | 31% |
| Other | | 5 | 13% |
| Joint research without Funding | | 26 | 67% |

*Table 3. Type of interactions industry has with academia.*

The next question asked industry if your organization seeks to hire PhDs, how would you characterize the current hiring climate? Respondents were asked to select one.

A majority of the respondents said that it is somewhat challenging to hire good PhDs for their positions. Other respondents elaborated on that point and said that it is hard to hire good PhDs because the PhD market is currently very competitive with top universities having slots to hire strong candidates. Big well known companies, Google, Facebook, and LinkedIn, offer extremely generous packages for new PhD's that make competing against them very challenging. This competition with the well-known companies is currently the biggest worry for many of the respondents (Table 5).

The final question of the industry survey asked the respondents, what value do you see in hiring a fresh PhD compared to someone with a master's degree and some years of experience? What aspects of PhD training would enhance this value for you? A majority of respondents said that PhDs were more valuable than someone with a master's degree and some years of experience because they can adapt, understand critical thinking, and have more expertise and independence in solving research problems (88%). Other respondents said that there was no difference between PhD students and master's students with work experience. A few even said that they prefer master's students with work experience rather than a fresh PhD student.

## Summary

There is a lack of communication and understanding between academia and industry. Industry hires undergrads and recent PhDs from academia. Still, there is a lot of mistrust of academics among those in industry (e.g., academics are not bound by time, don't care about end result, etc.). Both academia and industry struggle with understanding and agreeing on intellectual property. Intellectual property becomes an institutional issue on both sides and there is little room for maneuver. Both sides, however, seem to be open to collaboration and would love to see stronger relationships. There should be more initiatives for active collaboration between industry and academia. Having a way for industry to share their knowledge with academia is valuable, and vice versa. Collaboration on projects is the best way to accelerate both fields. Industry brings resources and scale. Academia has a high tolerance for risk. Together they could potentially take on very difficult problems and have tremendous success.

| Answer | Response | % |
|---|---|---|
| People-oriented — e.g., providing better methods or best practices for interaction/exchange of personnel between academia and industry | 14 | 41% |
| Process-oriented — e.g., creating better vehicles for exposing/engaging academic research programs with industry | 13 | 38% |
| Resource-related --- e.g., creating better mechanisms for shared data or infrastructure. | 14 | 41% |
| Communication-related — e.g., creating a clearing house for CS PhD students who would be interested in summer internships at a company. | 10 | 29% |
| More opportunities for people working in industry to attend, speak at, or publish at research conferences and journals? (e.g., industry tracks at conferences, conferences located near major cities or industrial hubs, etc.) | 13 | 38% |
| Better training of students for work in an industrial setting (e.g., professional programming practices, working effectively in teams, awareness of new technologies, good communication skills, etc.) | 15 | 44% |
| Other: | 7 | 21% |

*Table 4. Most effective ways to improve the connection between academia and industry.*

| Answer | Response | % |
|---|---|---|
| Fairly easy to hire good PhDs for our positions. | 2 | 6% |
| Somewhat challenging to hire good PhDs for our positions. | 23 | 64% |
| Very difficult to hire PhDs for our positions. | 5 | 14% |
| Any additional comments or explanations are welcome: | 6 | 17% |
| Total | 36 | 100% |

*Table 5. Current hiring climate.*

## Participants

Andy Bernat, CRA
Nady Boules, NB Motors
Khari Douglas, CCC
Ann Drobnis, CCC
Joel Emer, NVIDIA, MIT
Stuart Feldman, Google, Retired
Limor Fix, Intel, Retired
Michael Franklin, UC Berkeley
Greg Hager, JHU
Brent Hailpern, IBM Research
Peter Harsha, CRA
Laura Hass, IBM Research
Martial Hebert, CMU
University of Wisconsin- Madison
Alex Kass, Accenture Technology Labs
David Kriegman, Dropbox and UCSD
Sanjeev Kumar, Facebook
Jia Li, Snapchat
Arnold Lund, GE Global Research
Beth Mynatt, Georgia Tech
Klara Nahrstedt, UIUC
Christopher Re, Stanford
Chris Rossbach, VMWare
Shashi Shekhar, University of Minnesota
Maarten Sierhuis, Nissan Research
Stewart Tansley, Facebook
Min Wang, Visa
Helen Wright, CCC
Ben Zorn, Microsoft Research

This material is based upon work supported by the National Science Foundation under Grant No. 1019343. Any opinions, findings, and conclusions or recommendations expressed in this material are those of the authors and do not necessarily reflect the views of the National Science Foundation.

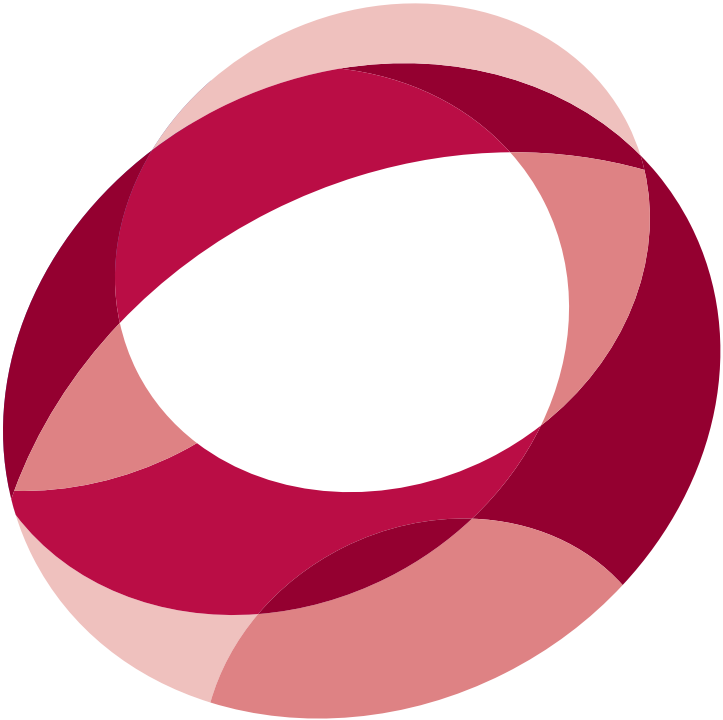